# Novel method for planar microstrip antenna matching impedance

Mahdi Ali, Abdennacer Kachouri and Mounir Samet

**Abstract—** Because all microstrip antennas have to be matched to the standard generator impedance or load, the input impedance matching method for antenna is particularly important. In this paper a new methodology in achieving matching impedance of a planar microstrip antenna for wireless application is described. The method is based on embedding an Interdigital capacitor. The fine results obtained by using a microstrip Interdigital capacitor for matching antenna impedance led to an efficient method to improve array antenna performance. In fact, a substantial saving on the whole surfaces as well as enhancement of the gain, the directivity and the power radiated was achieved.

**Index Terms—:** matching impedance. Interdigital capacitor. Planar microstrip antenna.

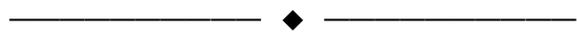

## 1 INTRODUCTION

In recent years, microstrip antenna has been more and more used in wireless equipment such as wireless sensors, RFID tags and cellular phones. Planar patch antennas are recognized to be the most useful type. However, patch antenna also have some limitations; narrow band width, large ohmic loss in the feed structure of arrays, reduced gain (6dB), reduced efficiency, complex feed structures required for high performance array [1].

Among the various approaches to enhance antenna effectiveness, novel feeding structures are proposed; balanced feed for RFID applications [2], balun structures [3], proximity electromagnetically coupled microstrip feed [1].

Basically, it is important to have an efficient input impedance matching the antenna with the load, to obtain maximum radiated power, many methods have been in use: stub [4], progressive balun [4], combined effects of the insert microstrip line and the slits [5].

This study is carried out to evaluate the efficiency of using a microstrip Interdigital capacitor to guarantee both feeding and matching impedance. Inter digital capacitors find application in filter, hybrid couplers, Dc blocking circuits, tuning impedance matching network.

In order to analyse the efficient results provided by using the microstrip Interdigital capacitor, we carried out a comparative study as well as a use of this structure for array antenna.

## 2 METHOD AND MATERIAL

For planar antennas structures, the Method of Moments or that of Finite Element are quite popular. In order to streamline the antenna design process and generate accurate results before prototype construction, it is important to select an EM simulation program. The soft used has been MOMENTUM which is one of the tools in Advanced design System 2005(Agilent).

Currently, a microstrip antenna consists of a dielectric substrate sandwiched between two conducting surfaces: the antenna plane and the ground plane. The simplified microstrip patch antenna is shown in Fig.1 For this study an epoxy dielectric was used (h=1,52mm, $\varepsilon_r$ = 4,32, metallisation layer thickness=35µm and substrate loss Tan$\delta$=0,018), patch dimensions are chosen such that the antenna resonates on a desired frequency (W=35and L=29mm resonate on 2.45 GHz).

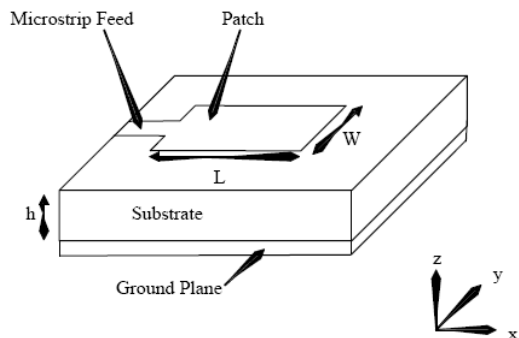

Fig. 1 Microstrip patch antenna construction

### 2.1 Approach

Basically, a matching juncture device is needed between the load and the feeding point in the patch antenna. This structure is shown in Fig.2.

_______________________

- *Mahdi Ali, Laboratoire d'Electronique et des Technologies de l'Information (LETI), Ecole Nationale d'Ingénieurs de Sfax.*
- *Abdennacer Kachouri, Laboratoire d'Electronique et des Technologies de l'Information (LETI), Ecole Nationale d'Ingénieurs de Sfax.*
- *Mounir Samet, Laboratoire d'Electronique et des Technologies de l'Information (LETI), Ecole Nationale d'Ingénieurs de Sfax.*

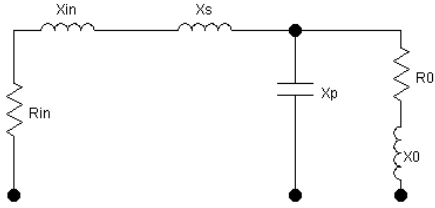

Fig. 2 Impedance matching solution

(Rin , Xin) the load impedance, (XS , XP) matching impedance and (R0 ,X0) antenna impedance.

Several publications have surveyed many possible types of microstrip antenna feeding, among which are the microstrip coplanar feed[6], aperture-coupled microstrip feed[7], proximity-coupled feed[8] and gap–coupled feed[9]. Typically a gap-coupled feed is shown in Fig. 3. It is worth noticing that a narrow gap width provides an efficient coupling of power.

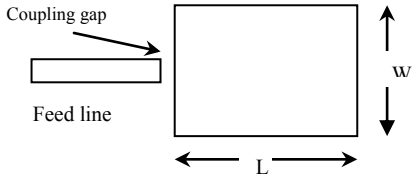

Fig. 3 Gap coupled patch

Gap introduces capacitive and coupling effects as the Fig.4 represents an equivalent lamped element circuit of a patch antenna fed by a strip line matched with a gap.

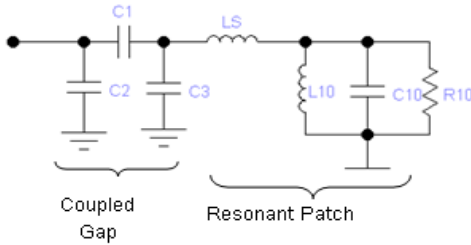

Fig. 4  Equivalent circuit patch fed with a line coupled gap

In order to enable tuning capacitive effect, an Interdigital capacitor can be introduced between the feeding device and the patch antenna not just for coupling but also for matching impedance. Fig.5 illustrates simulation of an antenna using the latter technique.

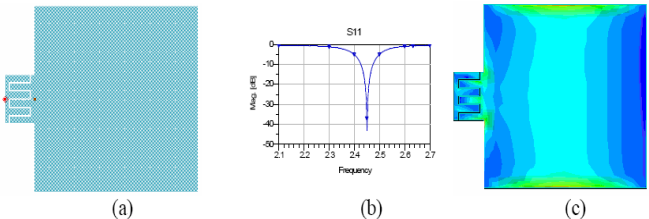

Fig. 5 Planar patch antenna using an Interdigital capacitor

Fig. 5 (a) shows a designed antenna fed with an Interdigital capacitor (b) the return loss S11 simulation results (c) current distribution.

A planar microstrip antenna fed with an Interdigital capacitor could be modulated with lamped elements circuit as illustrated in the Fig. 6

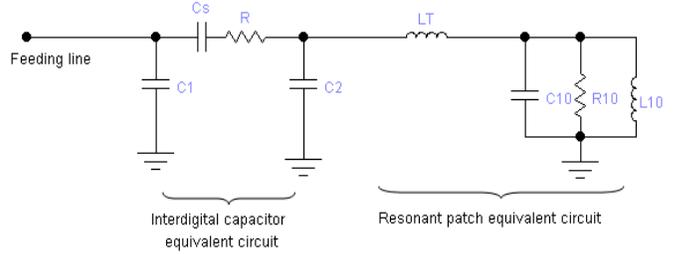

Fig. 6 Lamped–element equivalent circuit of a planar patch antenna

## 2.2 Patch antenna characteristics and model

### PATCH DIMENSIONS

The dimensions of the patch are expressed by the following equations:

$$W = \frac{c}{2f_0} \sqrt{\frac{2}{\epsilon_r + 1}} \qquad (1)$$

$$L = \frac{c}{2f_0 \sqrt{\epsilon_{eff}}} - 2\Delta L \qquad (2)$$

$$L_{eff} = L + 2\Delta L \qquad (3)$$

$$\Delta L = 0.412 \, h \left[ \frac{(\epsilon_{eff} + 0.3)\left(\frac{W}{h} + 0.264\right)}{(\epsilon_{eff} - 0.258)\left(\frac{W}{h} + 0.813\right)} \right] \qquad (4)$$

$f_0$ : Central resonance frequency taken 2.45 GHz .
The previous equations (1), (2), (3) and (4) gave W=37.5 and L=32, an adjustment is operated when simulating antenna leads to L=28,95 mm and  W= 35mm.

### INPUT IMPEDANCE OF A RECTANGULAR PATCH

The antenna has a physical structure derived from micro strip transmission line, the microstrip antenna is modeled as a length of transmission line of characteristic impedance[10] Z0(Ω) given by (5)(6).

$$Zo = \frac{120\pi}{2\sqrt{2\pi}\sqrt{\epsilon_r + 1}} \ln\left\{ \frac{4h}{w'} \left[ \frac{14 + {}^8/_{\epsilon_r}}{11} \frac{4h}{w'} + A \right] \right\} \qquad (5)$$

$$A = \sqrt{\left( \frac{14 + {}^8/_{\epsilon_r}}{11} \right)^2 \left( \frac{4h}{w'} \right)^2 + \frac{1 + {}^1/_{\epsilon_r}}{2} \pi^2} \qquad (6)$$

Were : $w' = w + \Delta w'$ and $\Delta w' = \Delta w \left( \frac{1 + 1/\epsilon_r}{2} \right)$      (7)

$$\frac{\Delta w}{t} = \frac{1}{\pi} \ln \left[ \frac{4e}{\left( \frac{t}{h} \right)^2 + \left( \frac{1/\pi}{w/t + 1.1} \right)^2} \right]$$      (8)

If we replace $\epsilon_r$ with $\epsilon_{eff}$

$Z_0 = \frac{60}{\sqrt{\epsilon_{eff}}} \ln \left( \frac{8h}{w} + \frac{w}{4h} \right) \Omega$ if $\frac{w}{h} < 1$      (9)

Otherwise

$Z_0 = \frac{120\pi}{\sqrt{\epsilon_{eff}}} \frac{1}{\left( w/h + 1.393 + 0.677 \ln(w/h + 1.444) \right)} \Omega$ (10)

$\epsilon_{eff}$ is expressed as follow

$\epsilon_{eff} = \frac{\epsilon_r}{\sqrt{1 + \frac{12h}{w}}} + 0.04 \left( 1 - \frac{w}{h} \right)^2$   if $\frac{w}{h} < 1$      (11)

Otherwise

$$\epsilon_{eff} = \left[ \frac{\epsilon_r + 1}{2} + \frac{\epsilon_r - 1}{2} \left[ \frac{1}{\sqrt{1 + 12h/w}} \right] \right]$$      (12)

Characteristic impedance evaluation of microstrip is important to determine the width of the feeding line calculated using (6).

The effective permittivity were calculated with (8) $\epsilon_{eff} = 4,022$.

The width of a $Z_0 = 50\Omega$ line is $w = 2,9438$ mm.

In order to evaluate the input impedance of a resonant rectangular patch antenna, a simplified model considering the most important mode TM10 proposed by [11] is showed in Fig.7 .

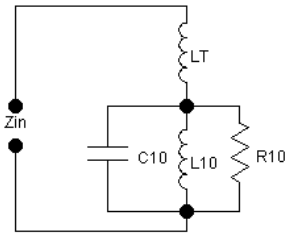

Fig. 7 Simplified equivalent lamped circuit of an antenna using resonant cavity theory [11]

Circuit element could be accurately calculated using empiric formulas.

On the other hand a simple analytical description of the rectangular planar patch antenna using transmission line model and models the patch as two parallel radiating slots[12] as shown in Fig 8 allow to evaluate the input impedance of the antenna.

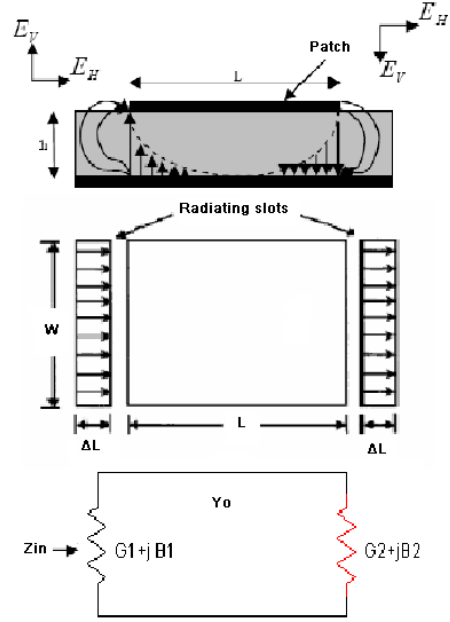

Fig. 8 The patch modelled as two parallel radiating slots having G1+jB1 and G2+jB2 as admittances

The slots admittance is given by (13).

$G_1 + jB_1 \cong \frac{\pi L}{\lambda_0 z_0} [1 + j(1 - 0.636 Ln(k_0 \Delta L))$      (13)

Where:

$\lambda 0$ is the free-space wavelength

$z0 = \sqrt{\mu_0 / \epsilon_0}$ and $k0 = 2\pi / \lambda_0$      (14)

$\Delta L$ is the slot width given by (17)

The slots are identical having the same admittance given by (13) accept for fringing effect effects associated with the feed point on edge 1[12].

For this typical design G1+ jB1= 0,00625+j0,0077

The input impedance is important to be evaluated. This parameter is important to determine the width of the microstrip feed line which allows the matching impedance between the patch and the load.

$Z_{in} = \frac{1}{2(G_1 + G_{12})}$      (15)

Where

$G_1 = \frac{1}{120\pi^2} \int_0^\pi C \sin^3 \theta \ d\theta$      (16)

$G_{12} = \frac{1}{120\pi^2} \int_0^\pi C \ J_0 \ (k_0 L \sin \theta) \sin^3 \theta \ d\theta$      (17)

Where $C = \left[ \frac{\sin\left( k_0 W/2 \cos \theta \right)}{\cos \theta} \right]^2$      (18)

J0 Bessel function of the first kind.

For this typical design Zin= 225Ω
In the case of feeding antenna with a quarter wavelength microstrip line, the line must have a width that satisfy a characteristic impedance Z0 calculated from (9)

$$Z_0 = \sqrt{Z_s \, Z_{in}} \qquad (19)$$

Where ZS is the impedance of the source and Zin is the input impedance of the patch.

## MODELLING AN INTERDIGITAL CAPACITOR

The role that the Interdigital capacitor is to match the real part of the input impedance of the patch, the capacitance is used to adjust the reactive component.

The properties of the Interdigital capacitor revealed on Fig.9, have been studied by many authors. We can mention that the capacitor dimensions are much less than a quarter wave length. Value of the capacitance depends on the number of fingers, gap wideness between the fingers, metallisation thickness[13].

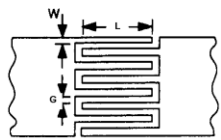

Fig. 9 A general form of the Interdigital capacitor [ADS]

An accurate study carried by [14] [15] established the lamped circuit equivalent of an Interdigital capacitor which is shown in Fig.10

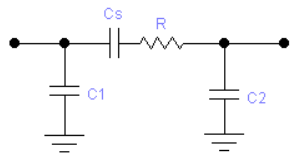

Fig. 10 Lamped–element equivalent circuit of an Interdigital capacitor

In this model, the series capacitor accounts for the capacitance between the fingers, whereas the series resistor represents metallization losses. The capacitors connected to ground are the parasitic capacitances.

Usually, R is very small and has a minor effect on the response of the Interdigital capacitor [14] on the other hand, the capacitance Cs is the most important element [16] presented the following expressions characterising the lamped element equivalent.

$$C_s = \frac{\epsilon_{eff}. \, 10^{-3} K(k)}{18 \, \pi \, K'(k)} (N-1) L \qquad (20)$$

K(k): is the complete elliptic integral of the first kind

$\dfrac{K(k)}{K'(k)}$ Could be approximated using the following expressions:

$$\frac{K(k)}{K'(k)} = \left( \frac{1}{\pi} \ Ln \left( 2 \frac{1 + \sqrt{k''}}{1 - \sqrt{k'}} \right) \right)^{-1} \quad 0 < k \le 0,7 \qquad (21)$$

Otherwise

$$\frac{K(k)}{K'(k)} = \frac{1}{\pi} \ Ln \left( 2 \frac{1 + \sqrt{k'}}{1 - \sqrt{k}} \right) \qquad 0,7 \le k \le 1 \qquad (22)$$

$$k = \tan \left( \frac{w \, \pi}{4 \, (w + G)} \right)^2$$

$$k' = \sqrt{1 - k^2} \qquad (23)$$

The series resistance of an interdigital capacitor can be calculated by using the following formula (20)

$$R_s = \frac{4 \, L}{3 \, w \, N} R_F \qquad (24)$$

Where RF is the frequency-dependent surface resistivity of the conductors

The previous expressions (16) and (20) are applied to choose the correct dimensions of Interdigital capacitor which need to be tuned once the design simulation is carried out.

## DESIGNING AND MEASURING ANTENNA EFFECTIVENESS

In order to provide evidence for the antenna effectiveness design, many parameters were measured. The efficiency of the matching device is proved by the minimum return loss S11, which influences the remnant characteristics such as Directivity, Gain and band width.

### Antenna designing:

The Fig.11 shows a photograph of the patch antenna matched with an Interdigital capacitor.

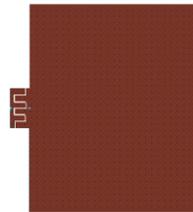

Fig. 11 Interdigital capacitor fed antenna

The antenna parameters are summarized in Table 1

TABLE 1
ANTENNA PARAMETERS

| Patch | W | 35mm |
|---|---|---|
| | L | 28.87mm |
| Substrate | εr | 4.32 |
| | h | 1.59mm |
| | Tanδ | 0.018 |
| metallisation | t | 0.0035mm |
| | Metal Permeability | 1 |
| | Metal Conductance | 1,83e+7 |

| Interdigital capacitor | Gap width | 0.1mm |
|---|---|---|
| | Finger large (w) | 1mm |
| | Finger length (L) | 1.717mm |
| | wt=wf | 0.64mm |
| | Nbr pair of finger | 3 |
| | Cs (pF) | 0.207 |
| | Rs (Ω) | 134 |

**Effectiveness evaluation**

The directivity is a measure of the directional properties of the antenna compared to those of an isotropic antenna. A simple approximate expression for the directivity D of a rectangular patch is given as formulated in [17] and [18]

$$D \approx \frac{4(k_0\,W)^2}{\pi\eta_0 G_r} \qquad (25)$$

$\eta$) = 120π Ω, Gr  is the radiation conductance of the patch.

The directive gain G of the antenna is defined as

$$G = e_r D \qquad (26)$$

where er  is the radiation efficiency of the antenna defined as the ratio of the radiated power Pr to the input power Pi ( 0<er<1).

# 3   RESULTS AND DISCUSSION

Validation of our analysis are demonstrated and discussed in this section defined as follow:

- Evaluating the accuracy of the matching impedance antenna resulting from simulating the return loss S11.
- Simulating the effectiveness and extracting the antenna parameters.
- Plotting currents.
- Simulating a conventional planar patch antenna and comparing results.
- Using the novel technique for array antenna.

## 3 .1 Return loss S11

Accurate matching impedance is provided by the adjustment of the Interdigital capacitor fingers number and length firstly obtained by using the analytic equation (18) the results of the simulation is given by the Fig.12.

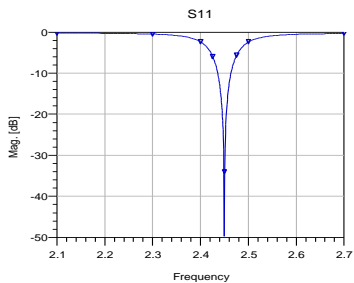

Fig. 12 Return loss S11 antenna simulation

The band width (-10dB) B =47MHz is deduced from the Fig.11

## 3. 2 Antenna parameters

Smith chart shown in Fig. 13 shows the input impedance for the antenna

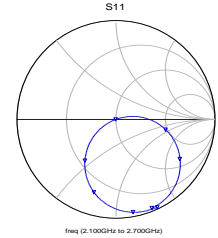

Fig. 13 Matching input impedance antenna

Table 2 shows the antenna parameter

TABLE. 2

ANTENNA PARAMETERS FED WITH AN INTERDIGITAL CAPACITOR

| Power radiated(watts) | 0.0409 |
|---|---|
| Effective angle(degrees) | 163 |
| Directivity(dB) | 6.4511 |
| Gain(dB) | 6.4501 |
| Max. Intensity(w/Steradian) | 0.0143 |

Thanks to a 2D data display, the effectiveness of the antenna is showed in Fig. 14, an efficiency of 99,9% was obtained.

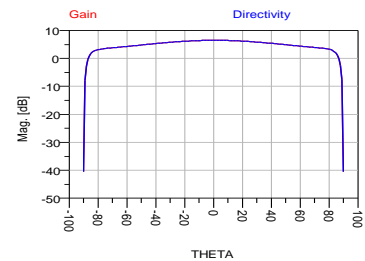

Fig. 14 Snap shut of the Gain and the Directivity of the antenna fed with Interdigital capacitor

Fig. 15 allows verifying the current distribution

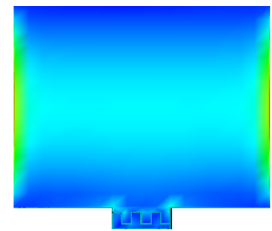

Fig. 15  Current distribution

The simulation of the same patch antenna designed later fed with a quarter wave length 50Ω line Fig.16 gave the performance shown in the TABLE.3

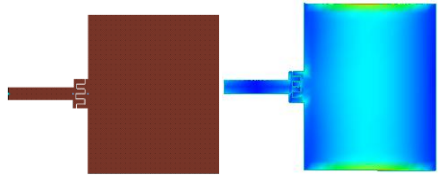

Fig. 16 simulation of a rectangular patch antenna fed with a quarter wave length 50 Ω line matched with an interdigital capacitor

TABLE.3

RESULTS OF THE ANTENNA FED
WITH MICROSTRIP LINE MATCHED WITH AN INTERDIGITAL
CAPACITOR

| Power radiated(watts) | 0.041 |
|---|---|
| Effective angle(degrees) | 162 |
| Directivity(dB) | 6.48 |
| Gain(dB) | 6.48 |
| Max. Intensity(w/steradian) | 0.0144 |

In order to evaluate these results, a comparison was carried out. A microstrip line fed patch antenna[19] Fig.17 is simulated using the same characteristics of the substrate previously used. The results are summarized in TABLE V The length line is quarter wave calculated using (27). The characteristic impedance of the line is 50Ω the width is obtained using (6) w=2,9438mm
In order to feed the patch at a 50Ω impedance point we operate slits.

$$\lambda_g = \frac{c}{f. \sqrt{\epsilon_r}} = 58,91mm \qquad (27)$$

The antenna dimensions are fully found in TABLE 4

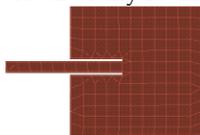

Fig. 17 Antenna with inset microstrip transmission-line feed

TABLE. 4

DIMENSIONS OF THE REFERENCE ANTENNA

| | | |
|---|---|---|
| Patch | W (mm) | 35 |
| | L (mm) | 30,22 |
| Substrate | $\epsilon_r$ | 4.32 |
| | H (mm) | 1.59 |
| | Tanδ | 0.018 |
| Feed line | Length (mm) | 14,72 |
| | Width (mm) | 2,9438 |
| | Slits length (mm) | 11,14 |
| | Slits width (mm) | 0,574 |

Simulation using ADS reveal the following results:

TABLE. 5

RESULTS OF THE ANTENNA FED WITH MICROSTRIP LINE

| Power radiated(watts) | 0.0343 |
|---|---|
| Effective angle(degrees) | 164 |
| Directivity(dB) | 6.413 |
| Gain(dB) | 5.754 |
| Max. Intensity(w/steradian) | 0.012 |

Current Distribution is obtained with a 3D simulation showed in Fig.18

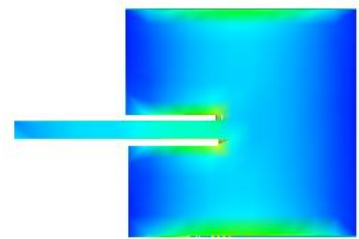

Fig. 18 Current distribution

The simulation of the antenna gave the return loss S11 shown on Fig.19

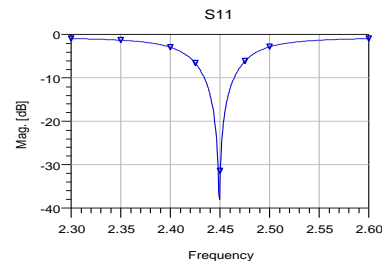

Fig.19 Return loss S11

The band width (-10dB) B =48 MHz is deduced from Fig.18
The efficiency of the antenna is plotted on Fig.20

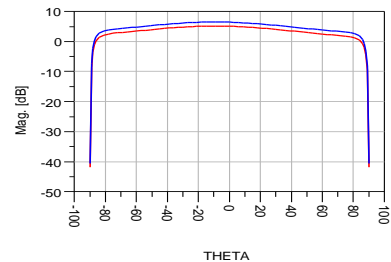

Fig. 20 Gain and directivity

The comparison between the results of the simulation of two patch antennas is summarized in TABLE 6

TABLE.6

COMPARISON BETWEEN INTERDIGITAL MATCHING AND LINE AND SLITS MATCHING

|  | Interdigital matching | slits | Enhancement % |
|---|---|---|---|
| Power radiated | 0.0409 | 0.0343 | +14 |
| Effective angle | 163 | 164 | - |
| Directivity(dB) | 6.4511 | 6.413 | - |
| Gain(dB) | 6.4501 | 5.754 | +10 |
| er  (%) | 99.9 | 89 | +11 |
| Max. Intensity | 0.014 | 0.012 | +17 |

The results obtained by using Interdigital capacitor may be efficient for array antenna. Fig. 21 shows a combination of two patches as an array antenna.

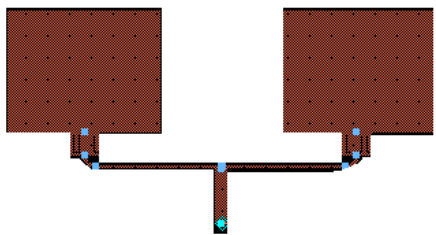

Fig. 21 Combination of two patches

The simulation of the antenna designed proved an effective distribution of current as demonstrated on Fig.22

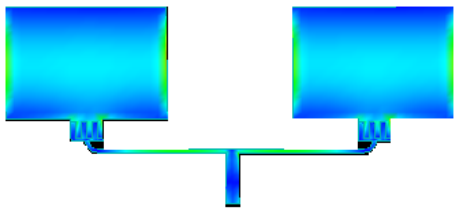

Fig.22 Current distribution

The performance of the array antenna are illustrated in TABLE.7

TABLE. 7

RESULTS OF COMBINATION OF TWO PATCHES

| Power radiated(watts) | 0.050165 |
|---|---|
| Effective angle(degrees) | 104.92 |
| Directivity(dB) | 8.373783 |
| Gain(dB) | 8.359342 |
| Max. Intensity(w/Steradian) | 0.027447 |

## 4  CONCLUSION

For the first time, an Interdigital capacitor has been used as matching impedance device for a planar microstrip antenna.

A comparative study was carried the results of which are summarized in TABLE 4. The Interdigital capacitor is useful as it allows the user to match the impedance of the antenna with any feed line. It is worth noticing the adjustment accuracy resulting from the modification of the finger length.

Moreover, the enhancement of the effectiveness of the antenna was proved; +14% of the power radiated, +11% er efficiency, +17% of the maximum intensity.

**Mahdi Ali** was born in Sfax, Tunisia, in 1962. He received the engineering diploma from National school of Engineering of Sfax in 2006, a Master degree in electronic from National school of Engineering of Sfax in 2008, a researcher in electronic in the "LETI" Laboratory ENIS Sfax.

**Abdennaceur Kachouri** was born in Sfax, Tunisia, in 1954. He received the engineering diploma from National school of Engineering of Sfax in 1981, a Master degree in Measurement and Instrumentation from National school of Bordeaux (ENSERB) of France in 1981, a Doctorate in Measurement and Instrumentation from ENSERB, in 1983. He "works" on several cooperation with communication research groups in Tunisia and France. Currently, he is Permanent Professor at ENIS School of Engineering and member in the "LETI" Laboratory ENIS Sfax.

**Mounir Samet** was born in Sfax, Tunisia in 1955. He obtained an Engineering Diploma from National school of Engineering of Sfax in 1981, a Master degree in Measurement and Instrumentation from National school of Bordeaux (ENSERB) of France in 1981, a Doctorate in Measurement and Instrumentation from ENSERB, in 1981 and the Habilitation Degree (Post Doctorate degree) in 1998. He "works" on several cooperation with medical research groups in Tunisia and France. Currently, he is Permanent Professor at ENIS School of Engineering and member in the "LETI" Laboratory ENIS Sfax.